\documentclass[varenna,nocopyright]{cimento}
\pdfoutput=1
\usepackage{graphicx}
\usepackage{amsmath}

\newcommand{\mb}{\mathbf}

\title{Quantum interferometry with complex molecules}

\author{Markus Arndt}
\institute{Faculty of Physics, University of Vienna,
Boltzmanngasse 5, 1090 Wien, Austria}

\author{Klaus Hornberger}
\institute{
Department of Physics, LMU Munich,
Theresienstra{\ss}e 37, 80333 M\"unchen, Germany}

\shortauthor{M. Arndt \atque K. Hornberger}

\PACSes{\PACSit{03.75.-b}{Matter waves}}

\begin{document}

\maketitle

\begin{abstract}
This chapter reviews recent experiments on matter
wave interferometry with large molecules. Starting from an
elementary introduction to matter wave physics we discuss
far-field diffraction and near-field interferometry with
thermally excited many-body systems. We describe the constraints imposed by decoherence and
dephasing effects, and present an outlook to the future challenges in macromolecule and cluster interferometry.
\end{abstract}

\section{Historical introduction and overview\label{history}}

The year 1923 is generally considered as the year of birth of
matter wave physics. It is the year when Louis de Broglie
proposed the idea that each material particle is accompanied by
an oscillatory phenomenon \cite{deBroglie1923a}. De Broglie based his hypothesis  on
two discoveries: the {\em energy-frequency equivalence} for
photons, $E=\hbar \omega$, and the {\em energy-mass equivalence}
for matter, $E= mc^2$. While the first relation had already been
central to Planck's explanation of the black body
spectrum~\cite{Planck1900a} and to Einstein's  analysis of the
photo effect~\cite{Einstein1905a}, the latter was a consequence of Einstein's theory of special relativity.

The generalization of these two relations to material particles
led Louis de Broglie to conclude that any object of mass $m$ must
be accompanied by an oscillatory phenomenon of frequency $\omega
= mc^2/\hbar$ \cite{deBroglie1923a}. If the material object moves
at velocity $v$ with regard to an observer at rest the Lorentz
transformation, which mixes the position and time coordinates,
implies that this oscillation appears as a propagating wave,
whose wave length is determined by the de Broglie relation
$\lambda_{\mathrm{dB}} = h/ (m v)$.

While this hypothesis could be immediately employed to explain the
stability of the electron orbits in Bohr's model of the hydrogen
atom, its formalization led to Schr\"{o}dinger's wave
equation~\cite{Schroedinger1926a}, one of the cornerstones of
modern physics. Empirical confirmations of the wave hypothesis
were soon found in two experiments with free electron beams: In
1927 C. J. Davisson and L. Germer described interference in the
{\em reflection} of an electron beam off a clean nickel crystal
surface~\cite{Davisson1927a}. In the same year, G.~P.~Thomson
\cite{Thomson1927a}  observed diffraction in the {\em
transmission} of a free electron beam through a thin sheet of
platinum.

Today, {\em electron interferometry} is an important element in the
surface sciences~\cite{Missiroli1982a}. { Electron microscopes}
explicitly exploit the fact that fast electrons are associated
with a de Broglie wavelength that can resolve surface structures
down to the level of single atoms. The coherence properties of
electrons can also be used to reveal the surface crystal structure
in low-energy electron diffraction ({\em
LEED})~\cite{Vanhove1979a}. Also holography, which was suggested
for electrons first~\cite{Gabor1948a}, is nowadays applied for
advanced phase imaging in the materials
sciences~\cite{Tonomura1987a}. The electron wave nature is
essential for the understanding of the working principle of the
scanning tunneling microscope ({\em STM}) \cite{Binnig1983a}, as
well as  for the interpretation of advanced STM
images~\cite{Crommie1993a}. Electron coherence is also the
precondition for decoherence experiments in mesoscopic
semiconductor structures~\cite{Ji2003a} and with charges flying
close to material surfaces~\cite{Sonnentag2007a,Gronniger2005a}.

As indicated by the de Broglie relation $\lambda_{\mathrm{dB}} =
h/ (m v)$, the observation of matter wave effects depends
crucially on the speed and mass of the interfering particles. The
smaller the mass and the velocity of the objects the easier it
will be to observe wave phenomena. In addition, neutral particles
are more accessible to interference compared to charged  ones,
since they are less susceptible to phase averaging and
decoherence induced by residual electromagnetic fields.

The lightest {\em neutral particles} that were experimentally
accessible at the beginning of the last century were the helium
atom and the hydrogen molecule. I. Estermann and O. Stern were
the first to demonstrate {diffraction of He and H$_2$} at a
crystal surface~\cite{Estermann1930a}, which confirmed the de
Broglie relation for composite objects. Two years after that the
neutron was discovered by J. Chadwick~\cite{Chadwick1932a}, and
already four years later the first successful  { neutron diffraction}
experiment was carried out~\cite{Halban1936a}.

Although it takes a nuclear reactor to generate an intense neutron
beam, the methods of {\em coherent and incoherent neutron
scattering}, pioneered by C. Shull~\cite{Shull1995a} and B.
Brockhouse~\cite{Brockhouse1995a}, are nowadays routine tools in
the field of condensed matter physics~\cite{Sears1989a}, as well
as for experiments on the foundations of quantum
optics~\cite{Rauch2000a}. They are particularly useful for the
structural analysis of materials which are composed of light
elements, such as carbon and hydrogen. These elements, which are
ubiquitous in all organic materials, are less accessible to x-ray
imaging due to their limited number of electrons, but they are
favorable candidates for neutron scattering since the masses of
the lattice atoms and the scattered neutrons are not too
different.

After the first demonstrations of atom diffraction it took nearly
60 years before {\em atom interferometry} was taken up again. The
new experiments were spurred by the progress in the development of
lasers and nano-fabrication methods as well as by  advances in
atom cooling techniques. In the mid-1980s, interference  of atoms
was demonstrated using standing laser light
waves~\cite{Gould1986a,Martin1988a}, nano-fabricated
gratings~\cite{Keith1988a}, and mechanical double
slits~\cite{Carnal1991a}. Extensions to complete multi-grating
atom interferometers became rapidly
available~\cite{Keith1991a,Kasevich1991a} and were applied to
various new problems. This lead to precision measurements of
atomic polarizabilities~\cite{Ekstrom1995a,Miffre2006c}, of the
Earth's gravitation~\cite{Kasevich1992a,Peters1999a} and
rotation~\cite{Riehle1991a,Gustavson1997a}, of the atomic recoil
during photon absorption~\cite{Weiss1993a}, and the index of
refraction for atoms passing through a dilute
gas~\cite{Schmiedmayer1995a}.

The extrapolation of matter wave interference to {\em diatomic
molecules} was then realized in 1994 for I$_2$ using optical
recoil gratings in a Ramsey-Bord\'{e}
configuration~\cite{Borde1994a}. In the same year, far-field
diffraction behind a nano-fabricated mask
led to the discovery of the weakly bound He-dimer~\cite{Schollkopf1994a}, and the MIT group extended their Mach-Zehnder interferometer experiments from
sodium atoms to Na$_2$~\cite{Chapman1995a}.

In 1995 the experimental realization of {\em Bose-Einstein
condensates} then redirected a major part of atomic physics
research towards the exciting field of ultra-cold quantum
degenerate gases,  also covered by two Varenna summer schools by
now~\cite{Varenna1999a,Varenna2008a}. A plethora of matter-wave
phenomena was observed in these macroscopic quantum ensembles,
and we refer the reader to the recent reviews and monographs in
the field
\cite{Dalfovo1999a,Leggett2001a,Pethick2002a,Pitaevski2003a,Bloch2008a}.
It is noteworthy that the interference phenomena observed with
dilute Bose-Einstein condensates are single-atom effects, even
though the condensate order parameter may represent the matter
wave field of more than ten million atoms in a typical BEC
experiment. In spite of this macroscopic number, the wave length
associated with the solution of the Gross-Pitaevski equation is
determined by the mass of the individual atoms alone. As a matter
of fact, in all BEC experiments it is desired to isolate the atoms
in a dilute gas in order to avoid larger aggregations during the
cooling process.

The {purpose of our present contribution} is to introduce the
reader to the  coherent manipulation of {\em large molecules}, i.e.,
strongly bound ensembles of atoms  which may be as hot as
2000\,K. In these experiments, matter wave physics is extended
into a mass and complexity domain where many molecular properties
mimic features of bulk media rather than those of single atoms.
This opens a way to detailed studies of the gradual transition
from quantum phenomena to classical appearances, and it offers the
possibility to measure properties of big molecules in advanced
interferometric experiments.

\section{Elementary matter wave optics}\label{sec:elementary}

Before we turn to the experiments of molecular quantum optics, it
is worth emphasizing that many of the observations can be well
understood by the concepts of classical wave optics. Our starting
point is  the Schr\"{o}dinger equation since the energies and
velocities involved remain safely on the non-relativistic side of
physics. Let us first consider a stationary situation, as it is
often encountered in interferometry, where the time dependence of
both the beam source and the external potential can be ignored.
In this case, one can establish a formal equivalence between the
Schr\"{o}dinger equation on the one hand, and the (generalized)
Helmholtz equation for the propagation of light waves on the
other hand.

Starting from the Schr\"{o}dinger equation for a particle
subject to the potential $V(\mb{r})$,
\begin{equation}
i\hbar\frac{\partial \psi(\mb{r},t)}{\partial t} =
\left[ -\frac{\hbar^2}{2m}\nabla^2+V(\mb{r})
\right]\psi(\mb{r},t)\,,
\end{equation}
the separation Ansatz  $\psi(\mb{r},t)=e^{-i\omega t}\phi(\mb{r})$
directly leads to the Helmholtz form,
\begin{equation}\label{eq:Helmholz}
  \left[ \nabla^2 +k^2 n^2(\mb{r})\right] \phi(\mb{r})=0
\,.
\end{equation}
Here, $k=\sqrt{2 m E}/\hbar$ is the vacuum wave number and $n(\mb{r})= \sqrt{1-V(\mb{r})/E}$ the index of refraction. The latter depends on position
and energy in the generalized case, but it reduces to unity for a vanishing
potential.

To understand the elementary phenomena of matter wave optics we
can thus use the concepts and the intuition gained from the study
of light propagation. Static potentials can be viewed as producing
the index of refraction variations required to implement lenses,
gratings, and other optical elements \cite{Adams1994a}. Moreover,
since the wave length $\lambda_\text{dB}=2\pi/k$ is typically
much smaller than the scale of potential variations, one can
usually take the short wave limit, thus switching from wave
optics to ray optics when accounting for macroscopic external
fields.

\section{De Broglie interference with clusters and large molecules \label{LargeMolecules}}
\subsection{Does size, mass or internal complexity affect the wave behavior?}\label{sec:motivation}
As is well known, quantum physics is the best tested theory of nature.
The concept of matter waves, which  triggered its formulation,
has been quantitatively confirmed with electrons, neutrons,
atoms, and small molecules. One may therefore ask whether it is
really necessary to further explore the wave nature of larger objects.

In fact, one can give quite a number of motivations for pursuing
research in that direction. First of all, according to our
everyday experience  the superposition principle, which is
central to quantum mechanics,  does not  appear to show up in our
macroscopic life. For instance, we never find the position of a
macroscopic object to be in a delocalized state, colloquially
alluded to as `being in different places at the same time'. This
colloquialism is actually a misnomer since position measurements
have definite outcomes also in the microworld, while the
delocalization can only be inferred. Yet, it remains an open and
important scientific question whether quantum rules apply on all
size, mass, and complexity scales.

It is therefore legitimate to ask whether there could be an
objective transition between the physical laws governing the
quantum and the classical world, or whether the observed loss of quantum behavior is only
a result of the fact that a complex environment is coupled to the quantum states of interest~\cite{Zeh1973a,Zurek1991a}. This problem is also related
to the question whether the unitary quantum evolution provides a
complete description of all phenomena, or whether the random
incidence associated with a quantum measurement process must be
regarded as a fundamentally non-unitary element
\cite{Bassi2003a,Schlosshauer2005a}. A number of  approaches try
to extend quantum mechanics by   relating this non-unitary
element to quantum aspects of space time.  Some of these
proposals predict effects, which become visible in matter wave
interferometry once the delocalized object exceeds a certain
mass, although they agree with
standard quantum mechanics at smaller masses~\cite{Penrose1996a,Bassi2003a,Leggett2002a,
Wang2006a,Carlip2008a,Wang2008a}. Independently of
one's attitude towards such unconventional approaches, it is the
genuine task of physicists to explore the practical limits of our
present understanding of physical phenomena.

We have also discussed before that matter wave interferometry has
found numerous practical applications, as precise 'meters' for
crystal structures, surfaces or external fields. Molecule
interferometry can add to that, in particular as a tool for
studying {\em molecular properties}. This is a new perspective,
related to the rich internal structure of the particle, which
grows exponentially with the number of internal constituents.
Also {\em molecular lithography} may benefit from quantum
interferometry~\cite{Arndt2002b}. The prospects and limitations
of  atomic lithography as a nano-deposition technique have been
discussed in~\cite{Meschede2003a}. Although a typical molecular
beam will hardly reach the flux of the best possible atomic
sources, it is worth noting that single molecules--in contrast
to single atoms--can often already be regarded as functional
elements by themselves. A single molecule may perform a certain
task, such as acting as a logical element in molecular
electronics or as a single-photon emitter in quantum optics.

Finally, interferometry with large molecules adds a plethora of
new objects to the field of quantum optics. New questions and
effects will arise due to their rich internal structure and
nontrivial properties, which can be tailored to a large degree in
the process of chemical synthesis.

Molecular quantum optics is thus an interdisciplinary endeavor,
combining quantum physics, physical chemistry and the
nano-sciences. It is therefore an exciting and important task to
develop the molecular beam methods, coherent manipulation
schemes, efficient detectors, and the tricks to avoid dephasing
or decoherence for future de Broglie experiments with large
molecules and supermassive clusters.

\subsection{Where is the challenge?}
The difficulties encountered in molecule interferometry are quite
substantial. They arise from the conspiracy of various hard
facts, which can be summarized as follows:
\begin{itemize}
  \item The de Broglie wavelength of an object decreases in proportion to the increase
  in its mass. There is nothing one can do about
    this scaling since we explicitly want to study very massive objects. Given the currently existing molecular beam
    methods and the presence of Earth's gravity one has to work with de Broglie wavelengths in the range of 10\,fm to
    10\,pm. This is a very small length scale,  ranging between one thousandth and one millionth of the size of each interfering single cluster or molecule.
  \item In contrast to atoms,  molecules cannot be easily controlled, slowed, or collimated by
  laser beams. Their rich internal   structure, as well as rapid state changes on the time scale of
  sub-picoseconds (vibrations) and nanoseconds (electronic
  transitions),  strongly impedes the effective external control. Over the last years, several experimental and
  theoretical groups have started analyzing and tackling this
  problem. There is, in particular, substantial progress in the handling of
  small molecules with switched quasi-static electric fields~\cite{Bethlem1999a,Bochinski2003a,Heiner2007a} and also using cavity assisted laser scattering
  methods~\cite{Gangl2000a,Lev2007a}.  The controlled manipulation of large species
  is yet still an open challenge for several years to come.
  \item   Atoms can nowadays be detected with an efficiency close to unity, even selectively with regard to their internal state. In contrast to that, many different detection
  schemes   have to be explored for molecules, since each particular
    species responds differently to electron impact, laser excitation for fluorescence and ion collection, scanning probe techniques, and other conceivable methods.
    \item Interference experiments probing the coherence of matter waves must scrupulously avoid any perturbations by the environment. Complex molecules
    offer many additional channels for the interaction with the environment and they are therefore highly susceptible to any dephasing and decoherence effects.
    \item Large and thermally excited molecules often resemble small lumps of  condensed matter. One consequence of this is that each individual
    many-body system may often be regarded as carrying along  its own internal heat bath. This can determine the  likelihood for
    exchange events between the quantum system and its environment, and thus affect the molecular coherence properties.
    \item Complex, floppy molecules may undergo many and very
    different conformational state changes even while they pass
    the interferometer. Several electro-magnetic properties, for instance the electric polarizability or the dipole moment, will change
    accordingly. This, in turn, can affect both the molecular
    interaction with the diffraction elements as well as their
    probability to couple to external perturbations.
\end{itemize}

The matter wave interference experiments with large molecules
carried out in Vienna prove, in spite of all these
difficulties, that quantum coherence can be generated, maintained and
clearly revealed for objects composed of more than one hundred
covalently bound atoms even at internal temperatures between a
few hundred to 2000\,K, i.e., at temperatures ten orders of
magnitude beyond those required for Bose-Einstein condensation.
These experiments also allow us to quantitatively investigate the
influence of various decoherence mechanisms, some of which may
even be used in a positive sense, for measuring molecular
properties. We will describe the experiments, the required
theoretical concepts, and a number of applications in the
following sections.

\section{Quantum coherence experiments: Concepts and realizations}

\subsection{Far-field diffraction of  C$_{60}$ at a nanomechanical  grating}

\begin{figure}[tbp]
\begin{center}
\includegraphics[width=\columnwidth]{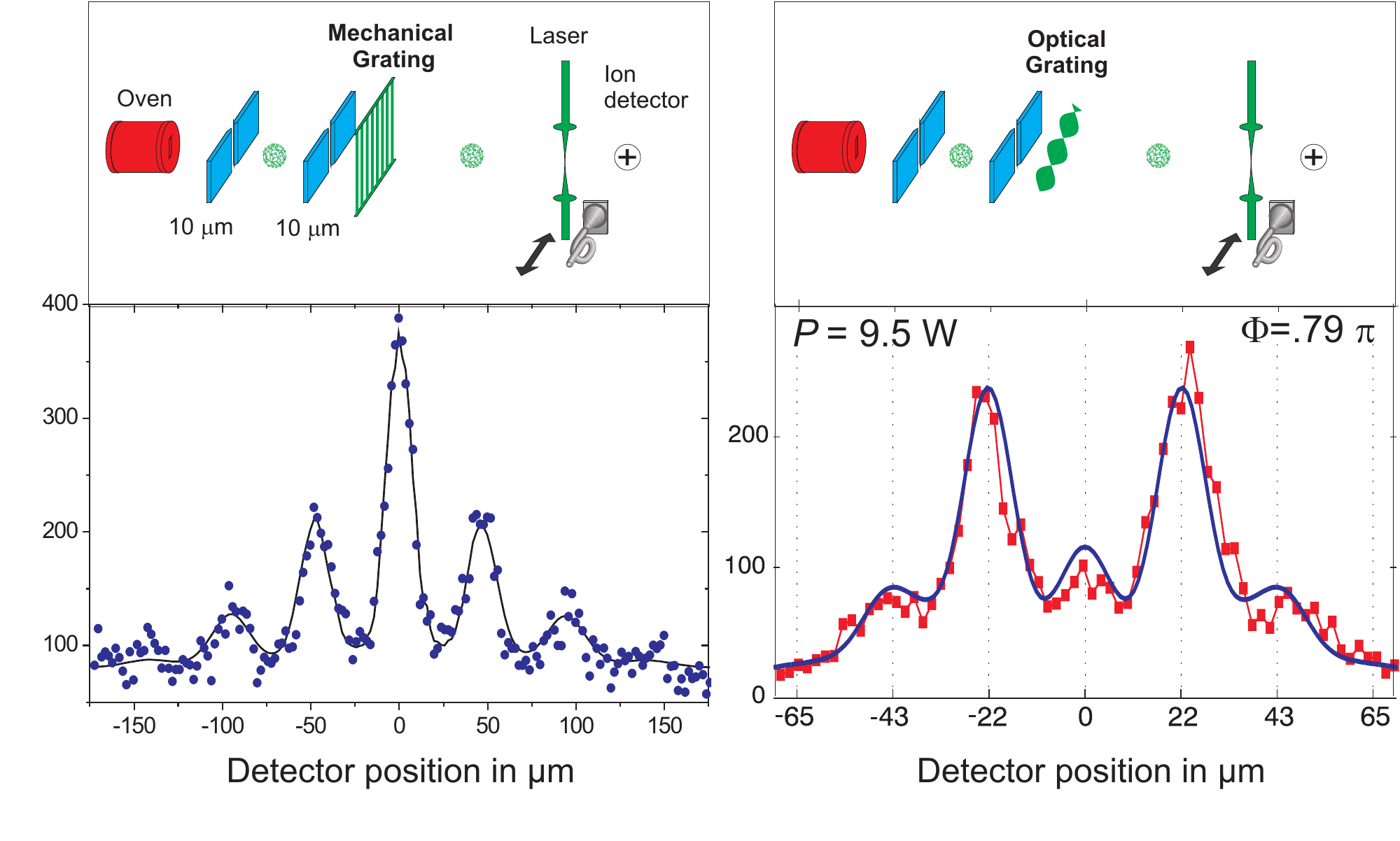}
\caption{Far-field experiments with C$_{60}$ fullerenes [64,66]. Upper left: setup
for diffraction at a nanomechanical grating. Upper right:
the mechanical grating is replaced by the standing light wave
of a green laser beam. Bottom row: the corresponding far-field interferograms
show a distinct difference in the center of the pattern [67].
Optical phase gratings
are better suited for creating a wider splitting between the
superposed molecular position states.} \label{fig1}
\end{center} \end{figure}

The coherent delocalization of large molecules was demonstrated
for the first time with the fullerene C$_{60}$, in a far-field
diffraction experiment using a nanomechanical diffraction grating
made of silicon nitride. The layout of the experiment is sketched
in the upper left part of Figure~\ref{fig1}. Fullerenes can be
sublimated in an oven at  a temperature of 900\,K. The
temperature is a compromise between a sufficiently large vapor
pressure and count rate on the one hand and the risk of thermal
fragmentation on the other hand. An effusive molecular beam is
created by a small orifice in the oven wall. At a temperature of
$T=900$\,K the most probable thermal velocity is $v=
\sqrt{2k_\text{B} T/m}=144$\,m/s, which corresponds to a de
Broglie wavelength of 3.8\,pm. This is already more than two
orders of magnitude smaller than the size of the molecule itself.

A central issue in all interference experiments is the question
of the transverse and longitudinal coherence. Since none of the
molecules in the beam knows anything about the phase of the
others we have to regard the molecule source in analogy to
thermal sunlight. Spatial or transverse coherence is prepared by
making sure that the source appears under a sufficiently
small solid angle when seen from the detector.  The collimation is
determined by the condition that the diffraction angle must be
larger than the collimation angle. The mechanical grating in
Fig.~\ref{fig1} has a period of 100\,nm. According to textbook
physics, we thus expect the first diffraction maximum at an angle
of $\theta_{\mathrm{diff}}=38\,\mu$rad. The collimation angle
should therefore be smaller than $20\,\mu$rad, which can be
realized with two slits of 10\,$\mu$m width, separated by about
1\,m.

The longitudinal coherence is determined by the velocity
distribution (spectral purity) of the beam and to a good
practical approximation described by the length
$L_c=\bar{\lambda}_\text{dB}^2/\Delta \lambda_\text{dB} $, where
$ \Delta\lambda_\text{dB} $ is the wavelength spread around the
mean value $\bar{\lambda}_\text{dB}$. A thermal molecular beam
has a typical velocity and wavelength spread of $\Delta
\lambda_\text{dB}/\bar{\lambda}_\text{dB} =60\,\%$. Without any
further measures, the coherence length is thus comparable to the
thermal de Broglie wavelength.   Using a rotating-disk velocity
selector, but also using gravitational selection schemes~\cite{Nairz2001a} the velocity bandwidth can be narrowed down to
16\,\%. The coherence length is thus increased to about six times
the mean de Broglie wavelength~\cite{Nairz2003a}, which is
sufficient to observe several higher diffraction orders in the
far-field behind the grating, as shown on the bottom left of
Fig.~\ref{fig1}.

Fullerenes were ideal starting candidates for macromolecule
interferometry since they can be prepared in high purity (better
than $99\%$). Moreover they have many similarities to small solid
state systems: At elevated temperatures they show thermal emission
of electrons, comparable to the thermionic emission of a hot
tungsten wire. They may emit molecular fragments, comparable to
the evaporation from a solid surface. Last but not least, the
photons emitted by hot fullerenes form a quasi-continuous
spectrum, strongly resembling that of a black
body~\cite{Kolodney1995a}.

Thermal ionization can be used in the detector stage: the
molecules are irradiated with a tightly focused green laser beam
which scans across the particle beam in a perpendicular
orientation. The generated ions are accelerated and counted in a
secondary electron multiplier as a function of the laser
displacement. The result of this experiment is shown in
Fig.~\ref{fig1}. This experiment represents the first observation
of quantum interference with a thermally highly excited many-body
system~\cite{Arndt1999a,Nairz2003a}.

These first studies revealed already that van der Waals forces
acting between the molecules and the grating walls imprint
position and velocity dependent phase shifts, and thus impose
stringent limits on the ability to observe matter wave
interference. An indication for this fact can be obtained from a
more detailed analysis of Fig.~\ref{fig1}. As we know from our
introductory optics classes, every far-field diffraction pattern
of a  grating with period $d$ is modulated by an envelope given
by the diffraction pattern corresponding to a single-slit. The
interference maxima behind a grating are found at integer
multiples of the angle $\theta \simeq \lambda_\text{dB}/d$,
whereas the minima of the single-slit envelope are positioned at
$\theta \simeq \lambda_\text{dB}/a$. Since our nanomechanical
grating was designed with a period of $d=100\,$nm and an open
slit width of $a=50\,$nm, the second order grating diffraction
peak should therefore be completely suppressed by the single slit
envelope.

The presence of the second order diffraction peak in
Fig.~\ref{fig1} is consistent with the assumption that the
attractive van der Waals interaction between the traversing
molecule and the grating wall reduces the effective slit width by
almost a factor of two. This huge effect inspired experiments
with optical phase gratings, where the electric field of a laser
creates a dipole potential proportional to the optical
polarizability of the molecule~\cite{Nairz2001a}. While phase
gratings were already known for atoms~\cite{Gould1986a}, the
internal molecular structure adds again to the complexity of the
diffraction physics, since the molecular line widths can be as
broad as 30\,nm to 50\,nm  at elevated temperatures. The
electromagnetic field therefore does not induce virtual
transitions corresponding to a single, spectrally sharp atomic
line, but rather acts on the far-detuned wings of all molecular
resonances, where the optical polarizability approaches the
static polarizability.

\subsection{Molecular far-field diffraction at a standing light grating}

Figure~\ref{fig1} shows on the right-hand side the experimental
configuration where a retro-reflected focused Argon ion laser
beam (with a wavelength of $\lambda_\text{L}=514\,$nm) forms a phase
grating with a period of 257\,nm for the passing C$_{60}$
molecules. The far-field diffraction pattern presented in the
lower right panel of Fig.~\ref{fig1} is in very good agreement
with the theoretical expectation.

Such phase gratings of light have several distinct advantages over
nano-fabricated structures:  Even lasers with a moderate
bandwidth produce nearly perfect periodic gratings which neither
clog nor break. An additional knob is offered in the experiment by the possibility to vary the laser intensity, and the problem of van der Waals
forces is eliminated at the expense of somewhat increased alignment
requirements. A new aspect is the possibility of photon
absorption, although the absorption cross section
is sufficiently small in the case of C$_{60}$  to be negligible for the experiments.

However, interference fringes can be observed also for C$_{70}$
fullerenes, where the cross section is larger by almost an order
of magnitude, so that on average one photon is absorbed during the
molecular passage through the grating. This highlights  a
peculiarity of large molecules: After absorbing a photon they
preferentially channel the photonic energy into vibrational
excitations rather than in re-emitting fluorescent light. This
intra-molecular energy increase occurs coherently at all positions
within the standing light grating, so that the internal and the
motional states of the molecule remain separable. At the same
time, the absorption of a single photon changes the transverse
momentum of the molecule by $h/\lambda_\text{L}$. The associated
far-field probability distribution then gets shifted by exactly half a period, which blurs the fringe pattern. In contrast, the absorption of two photons  hardly affects the visibility because it shifts by a full period (or not at all if the photon momenta have opposite momentum). Since the probability for absorbing no or two photons is always greater than the likelihood of a single photon absorption, a fringe pattern can still be observed. This is
different to the effect of photon scattering in atom
interferometry experiments~\cite{Pfau1994a,Chapman1995a}, where
the photon may be scattered in all directions, thus blurring the
interference fringes very efficiently.

\section{Near-field interferometers for massive molecules}\label{sec:TL}
\begin{figure}[tbp]
\begin{center} \includegraphics[width=0.8\columnwidth]{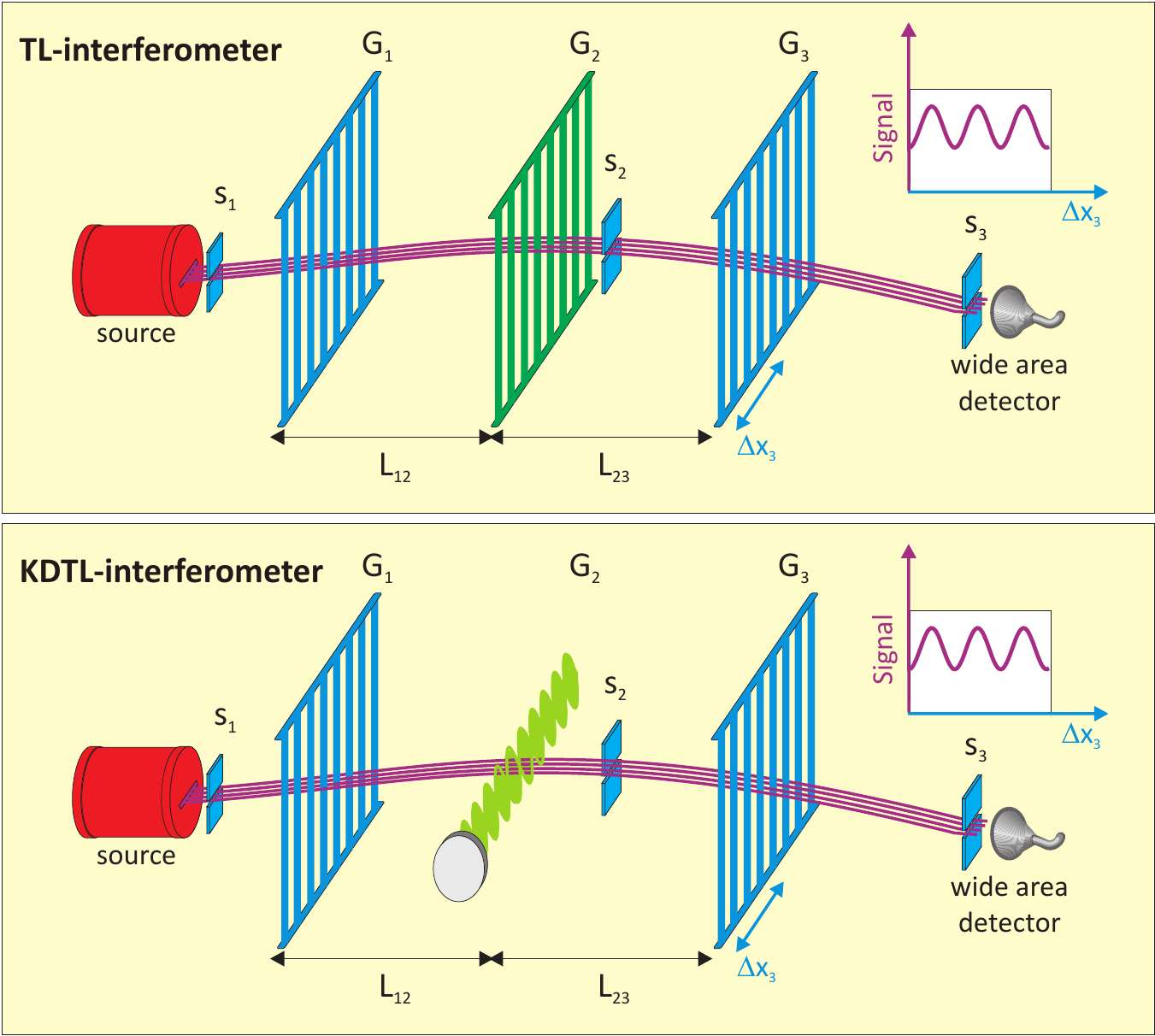}
\caption{Setup of the mechanical Talbot-Lau interferometer (top)
and the Kapitza-Dirac-Talbot-Lau interferometer (bottom). Top: The
gratings in the   TLI are made of gold, with a period of about
1$\mu$m, separated equidistantly by a distance between 22\,cm and 38\,cm, which
corresponds to one or two Talbot lengths, depending on the
molecular velocity. Bottom: The first and third grating in the KDTLI
are made of a 190\,nm-thick silicon nitride membrane with a
grating period of 257\,nm. This corresponds to half of the wave
length of the  laser beam, which is used to create a standing
light wave by retroreflection. In the experiment it is crucial
that the periods of the mechanical and the light gratings match exactly, already a deviation of
0.5{\AA} would substantially reduce the fringe visibility. }
\label{fig2}
\end{center}
\end{figure}

Observing a far-field diffraction pattern behind a material
grating probably constitutes the most direct prove of the
wave-nature of matter. However, as discussed above, this requires
the preparation of sufficient coherence, which in turn is based
on the tight collimation of the incident incoherent molecular
beam. Molecules which are even more massive than the fullerenes
will give rise to smaller diffraction angles and, correspondingly,
they will require even tighter collimation angles. We thus
encounter a very natural practical limitation for this type of
experiments below a de Broglie wavelength of around 1\,pm. In
combination with the 100\,nm sized diffraction structures, such
waves interfere constructively at angles close to 10\,$\mu$rad.
The  preparation of coherence would thus require the molecular
beam to be collimated below a few microradians, which borders on
the technologically possible. Of course, smaller grating
structures would yield wider diffraction angles, but the
aforementioned influence of van der Waals forces and, ultimately,
the lateral dimensions of large molecules set here strict limits,
too.

In order to demonstrate the wave nature of massive molecules it
is therefore highly desirable to work with wide grating openings
and less demanding collimation requirements. The solution is found
in near-field interferometry, which can be implemented in analogy
to the optical counterparts discussed by H. F.
Talbot~\cite{Talbot1836a} and E. Lau~\cite{Lau1948a}. This idea
was first implemented for atoms~\cite{Clauser1994a}, and also
suggested for high-mass experiments~\cite{Clauser1997a} by J.
Clauser. The first near-field interferometer for molecules was
then realized in our group in Vienna in 2002~\cite{Brezger2002a}.

The Talbot effect is a coherent self-imaging phenomenon. It
describes the fact that a plane wave traversing a periodic
absorptive structure will image this structure in integer
multiples of the Talbot distance $L_T=d^2/\lambda_\text{dB}$ by
virtue of the resonant near-field interference of different diffraction orders~\cite{Talbot1836a}.
Remarkably, the coherent self-imaging can be employed even for
spatially incoherent sources if one adds a second grating, as
shown in Fig.~\ref{fig2}. In this arrangement, each slit in the
first grating may be viewed as selecting an elementary spherical
Huygens wavelet, which propagates towards the second grating, thus
creating the required coherence. If the de Broglie wavelength
meets the Talbot resonance criterion, an interference pattern
appears behind the second grating as an overlay of all the single
self-images originating from the individual slits of the first
grating. In the simplest case, where the distance $L_{12}$
between the first and the second grating  equals the distance
$L_{23}$ between the second grating and the detection plane, the
image emerging in the plane of the third grating shows the
highest visibility provided that the period of the array of slit
sources G$_1$ equals the period of the second grating $G_2$. In
contrast to the pure Talbot effect, where self-imaging occurs precisely at
integer multiples of the Talbot-distance, the Talbot-Lau contrast
may be quite small at exactly this distance but it rapidly increases in the neighborhood if the grating separation is symmetrically stretched or
compressed [73,74].

\begin{figure}[tbp]
\begin{center}
\includegraphics[width=0.6\columnwidth]{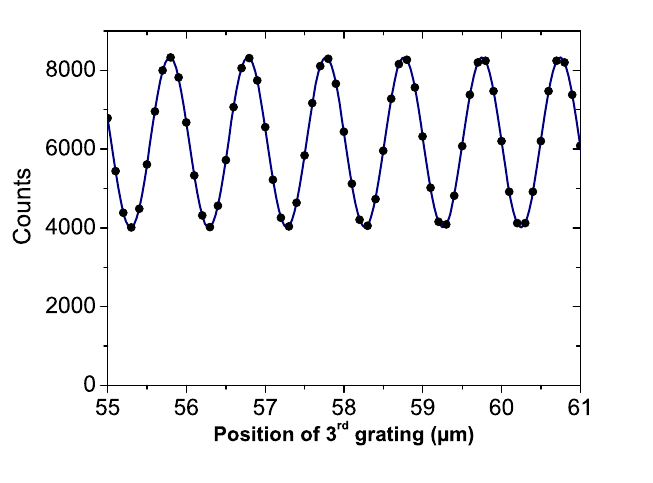}
\caption{Typical molecular density pattern at the exit of a
Talbot-Lau interferometer. The fringe period of 990\,nm is
determined by the period of the nanomachined grating. It is
therefore independent of the de Broglie wavelength. The fringe
visibility, however, is strongly dependent on the de Broglie
wavelength, which is set by choosing the beam velocity, see Fig.~\ref{fig4}.} \label{fig3}
\end{center} \end{figure}

There are many ways to detect the Talbot-Lau interferogram in the plane
$G_3$. A method that combines high spatial resolution with high
detection efficiency places a third grating in the image
plane, as indicated in Fig.~\ref{fig2}. The latter has the same period as the interferogram so that  the transmitted molecular flux gets modulated in its  intensity if we vary the transverse  position $x_s$ of the third grating. Plotting the total  molecular intensity $S(x_s)$ behind the third grating
as a function of its position $x_s$, directly reveals the
interference pattern. This 3-grating arrangement is designated as
a Talbot-Lau interferometer.
The first molecule interferometer exploiting this principle was
made from three gold gratings with a period of 990\,nm
\cite{Brezger2002a}. Figure~\ref{fig3} presents a typical
interferogram, recorded for the molecule tetraphenylporphyrin
displayed in Fig.~\ref{fig5}.

For our typical experimental conditions the interference pattern
is well described by a sine  curve, whose period is given by the
diffracting structure. Such a periodic curve is readily characterized by its contrast or visibility, defined in terms of the maximum and minimum count rates as
\[V=\frac{S_{\mathrm{max}}-S_{\mathrm{min}}}{S_{\mathrm{max}}+S_{\mathrm{min}}}\,.
\]
If the signal varies  by a factor of two, as is roughly the case in
Fig.~\ref{fig3}, its fringe visibility amounts to V=33\%.
\begin{figure}[tbp]
\begin{center}
\includegraphics[width=0.75\columnwidth]{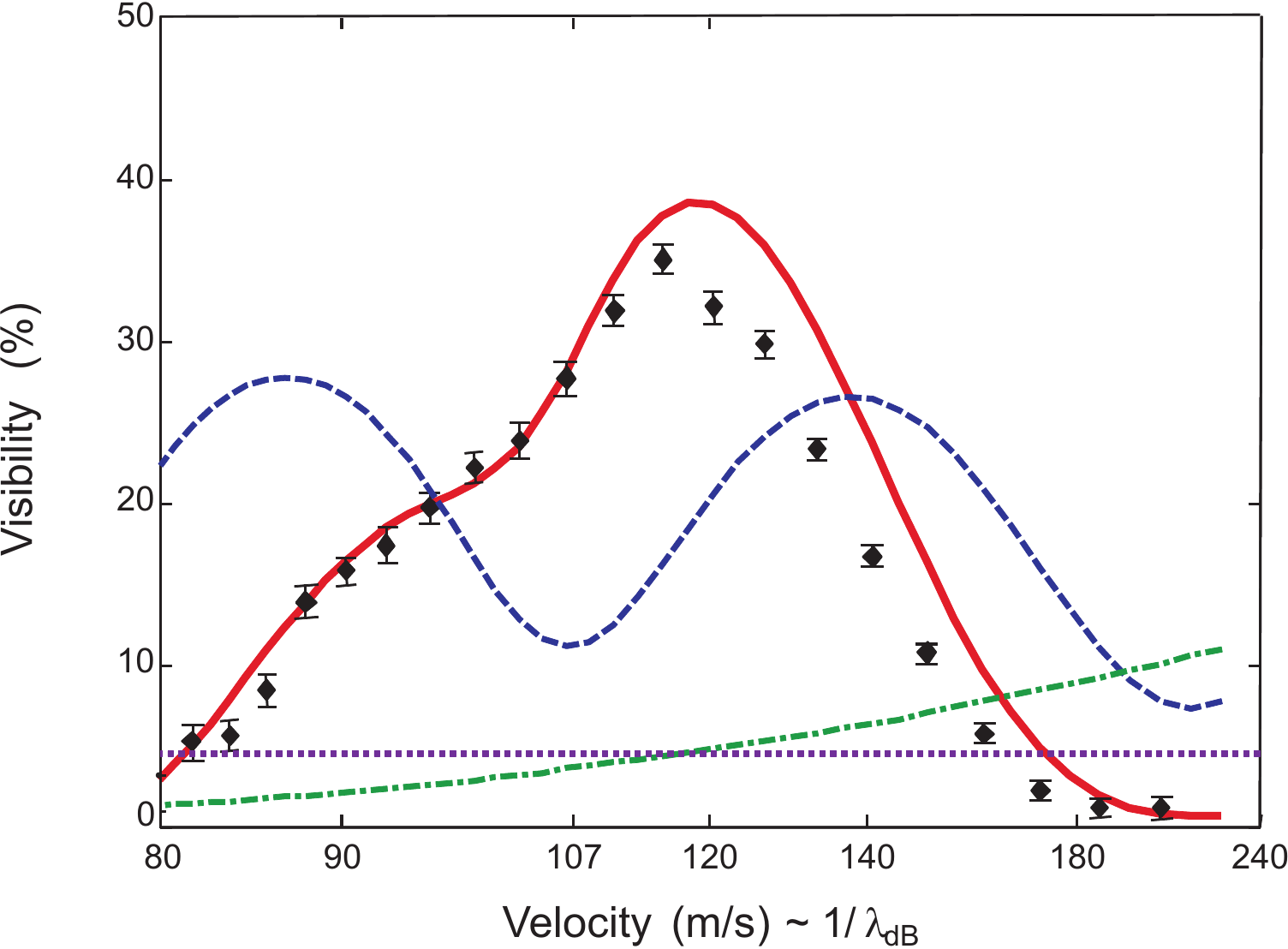}
\caption{Dependence of the fringe visibility of C$_{70}$
interferograms on the de Broglie wave length, as characterized by
the mean beam velocity. The full diamonds give the experimental
result obtained in a mechanical TL-interferometer, with a
molecular velocity spread $\Delta v/\bar{v}$ ranging from 8\% to
35\% at the large mean velocities \cite{Brezger2002a}. The
theoretical curves account for this distribution of wavelengths
in the beam. The dashed line gives the quantum mechanical
expectation, assuming that the molecules are ideal point
particles which do not interact with the grating walls. A good
agreement with the experimental observation is only obtained by
accounting for the forces exerted on the polarizable molecules
due to the van der Waals interaction with the grating walls
(solid line). The discrepancy at high velocity is reduced if one
one includes the effects of retardation as described by the
Casimir-Polder potential. The bottom lines are the result of a
classical calculation assuming that the molecules follow
Newtonian trajectories.  Unlike the dotted line, the dash-dotted
one includes the van der Waals forces. One observes that only the
quantum mechanical calculation is able to describe the experimental data. }
\label{fig4}
\end{center} \end{figure}
We can now use the fringe contrast to clearly distinguish genuine
quantum interference from a classical moir\'{e}-type shadow image,
which might also result from the classical flight of point particles through a
sequence of two or more gratings. As shown in Fig.~\ref{fig4}, the
velocity dependence (i.e., wavelength dependence) of the quantum fringe
visibility differs both quantitatively and qualitatively from all
classical predictions.

A detailed analysis~\cite{Hornberger2004a,Nimmrichter2008a} must
again include the influence of van der Waals forces between the
molecules and the grating walls. It turns out that the effect of
this dispersive interaction is  much stronger than in far-field
interference. This can be seen in Fig.~\ref{fig4}, where the
dashed line gives the quantum wave prediction for Talbot-Lau
interference of C$_{70}$ if we disregard the existence of van der
Waals forces. The experimental observation differs qualitatively, and only by accounting for the dispersive interaction we obtain a good
quantitative agreement with the experimental
observation~\cite{Brezger2002a}, as represented by the solid line.
An even better agreement is obtained in the high velocity wing if
the Casimir Polder potential, i.e, the fully retarded version of
the interaction is used (not shown here).

The Talbot-Lau interferometer was applied to demonstrate the
wave-like behavior of particles as massive as C$_{60}$F$_{48}$, cf Fig.~\ref{fig5}b, which comprises already more than one hundred atoms in a round
compound of 1632\,amu~\cite{Hackermuller2003a}. In the same setup we observed high-contrast
interference fringes of tetraphenylporphyrin
C$_{44}$H$_{30}$N$_4$~\cite{Hackermuller2003a}, an extended and
strongly oblate biodye, which is a relative to the color centers
in hemoglobine and chlorophyll, see Fig.~\ref{fig5}c.

Attempts to go to even larger molecules in this particular
experiment are impeded by the increasingly strong effect of the
dispersive interaction with the grating walls~\cite{Gerlich2007a}.
For molecules with a similar composition, the electric
polarizability increases with the particle size, and also the
grating interaction time increases with the mass, since smaller
velocities are required to keep the diffraction angles and thus
the Talbot length constant. The grating interaction thus imposes
increasingly strict requirements on the longitudinal coherence in
the molecular beam, which becomes rather demanding beyond masses
of 1000\,amu.

\begin{figure}[tbp]
\begin{center} \includegraphics[width=0.9\columnwidth]{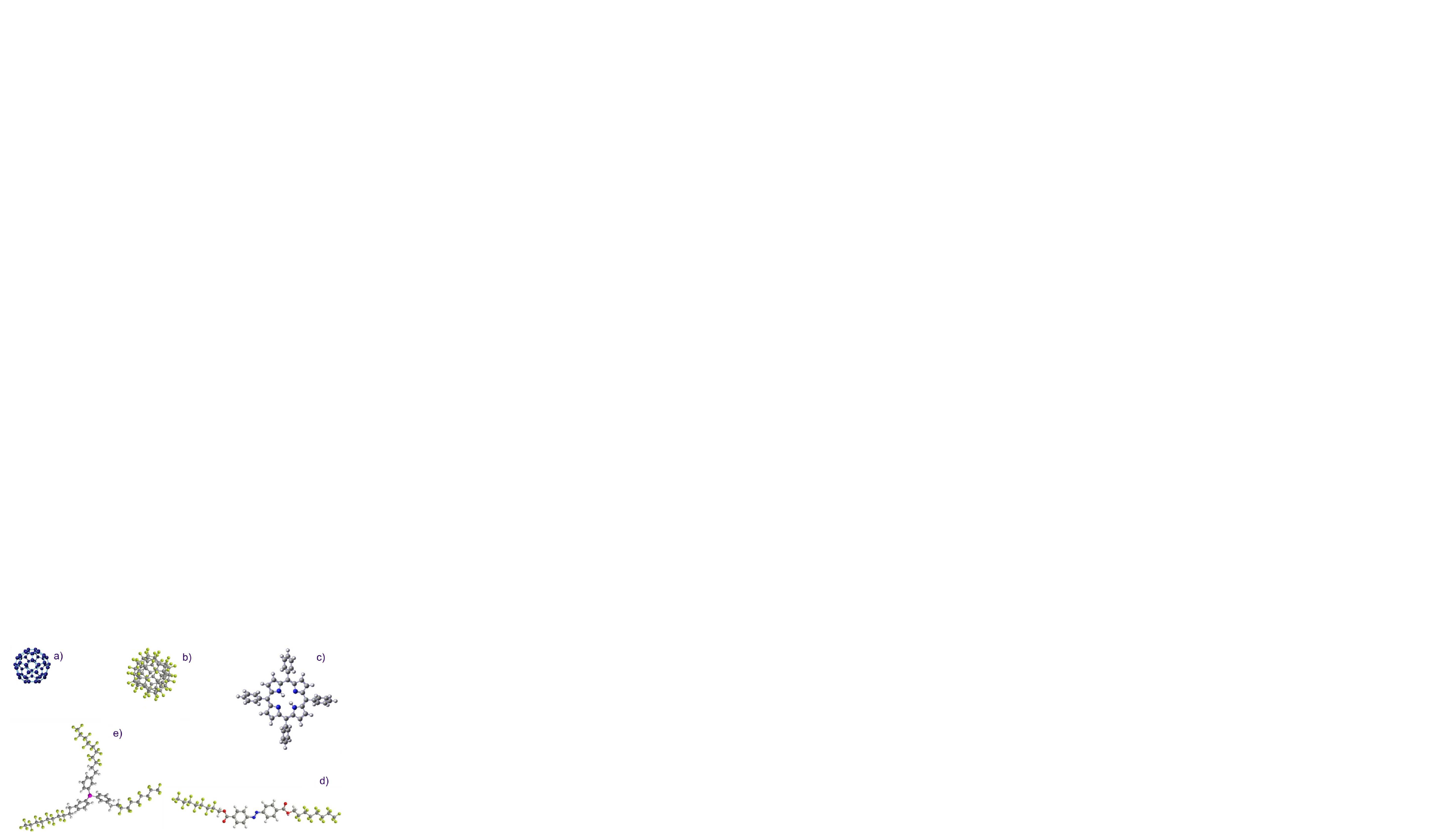}
\caption{Gallery of molecules that revealed their quantum nature
in the Viennese Talbot-Lau setup and in the
Kapitza-Dirac-Talbot-Lau interferometer.
 The relative size of the
molecules is roughly to scale. a) C$_{60}$ buckyball
~\cite{Arndt1999a}. b) fluorinated fullerene
C$_{60}$F$_{48}$~\cite{Hackermuller2003a}. c) Tetraphenylporphyrin
C$_{44}$H$_{30}$N$_4$~\cite{Hackermuller2003a}. d)
perfluoroalkyl-functionalized diazobenzene
C$_{30}$H$_{12}$F$_{30}$N$_2$O$_4$ \cite{Gerlich2007a}. e)
C$_{48}$H$_{24}$F$_{51}$P, a fluorinated catalyst molecule
\cite{Gerlich2008a}.} \label{biodye}\label{fig5}
\end{center} \end{figure}

In order to circumvent these dispersive interaction effects, we
replaced the central material grating by an optical phase grating,
as shown in the lower panel of Fig.~\ref{fig2}. We call this
setup a Kapitza-Dirac-Talbot-Lau
interferometer~\cite{Gerlich2007a}, since it combines the idea of
coherent self-imaging, as present in the Talbot-Lau concept, with
diffraction of matter at standing laser light gratings, as
originally proposed by Kapitza and Dirac for the diffraction of
electrons~\cite{Kapitza1933a}. The phase shift imprinted by the
standing light wave can be easily adjusted by the laser power.

Using the  the Kapitza-Dirac-Talbot-Lau interferometer we
demonstrated the wave nature of perfluoroalkyl-functionalized
azobenze molecules \cite{Gerlich2007a}. These polyatomic
molecular chains have a length of 32 {\AA}. In their stretched
trans-conformation they are four times more extended than the
fullerene C$_{60}$. In comparison to the soccer ball shaped
fullerenes, the azobenzene derivatives rather resemble strings,
and numerical simulations~\cite{Doltsinis2008a} confirm that they
undergo lively configurational state changes on a time scale
shorter than the passage time through the interferometer. In
spite of that complicated internal dynamics, they produce
interference fringes with a contrast which is in full
quantitative agreement with a quantum calculation based on its
molecular mass and static scalar polarizability. Similarly, we
observed undiminished interference of C$_{48}$H$_{24}$F$_{51}$P,
a fluorinated catalyst molecule with a mass of 1600\,amu which is
comparable to C$_{60}$F$_{48}$, but significantly more extended
~\cite{Gerlich2008a}; its three-legged configuration is shown in
Fig.~\ref{fig5}e.

\section{Decoherence and dephasing in matter wave interferometry}
\label{sec:decoherence}
\subsection{The concept of decoherence}

\begin{figure}[tbh]
\begin{center} \includegraphics[width=0.9\columnwidth,clip=true,trim=2cm 2cm 2cm 10cm]{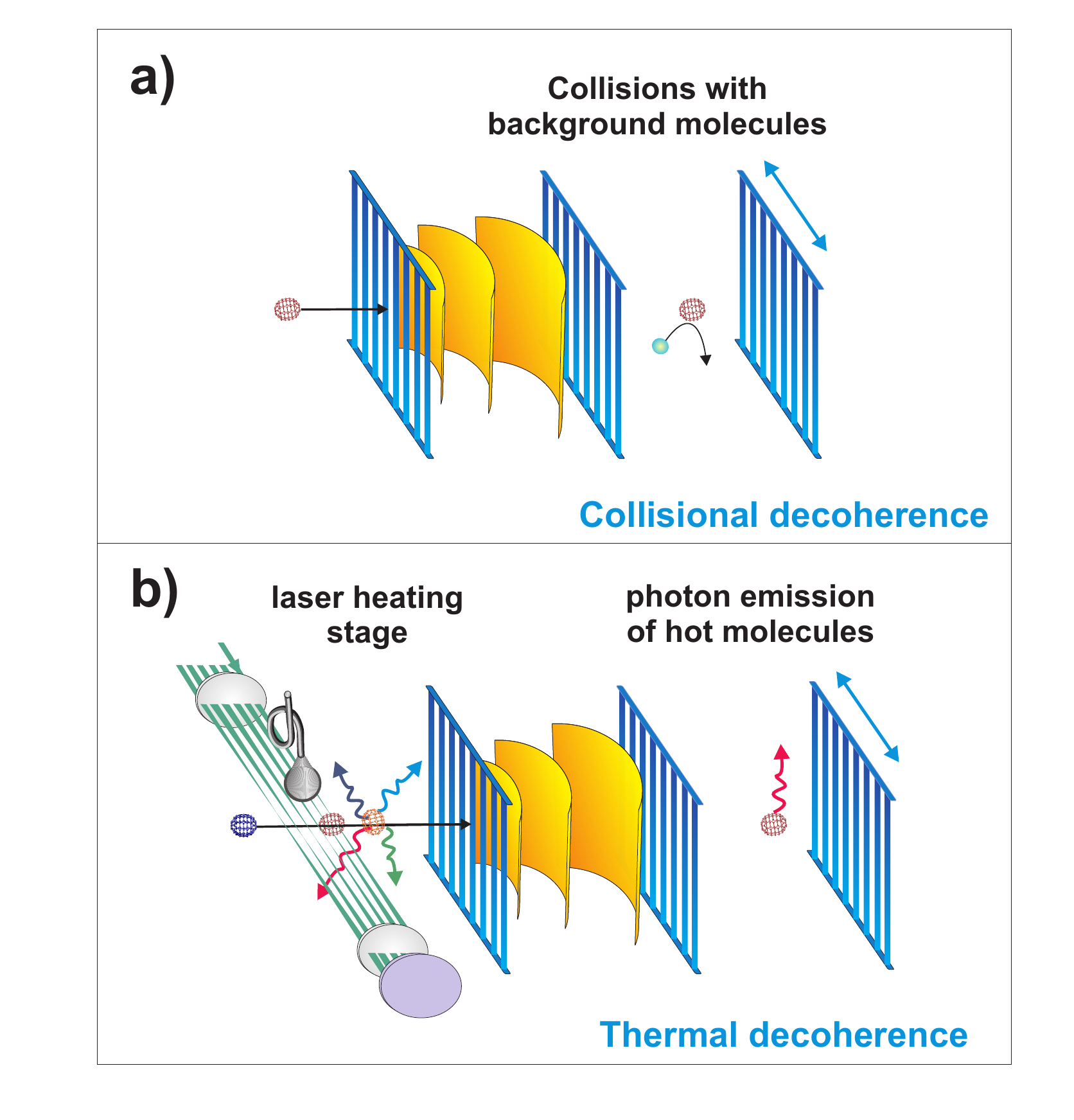}
\caption{Decoherence is a consequence of the  coupling of the
quantum system to a practically uncontrollable  environment. In
our  experiments we studied (a) the effect of collisions with
residual background gases \cite{Hornberger2003a} and (b) the
localization of hot fullerenes through emission of heat
radiation~\cite{Hackermuller2004a,Arndt2005b}.} \label{fig6}
\end{center}
\end{figure}

The  experiments described so far clearly indicate that even
large and complex molecules can exist in the delocalized state
required for interference at a grating. This is quite remarkable
since such molecules are usually viewed as well-localized objects that we can
even observe in high-resolution microscopy.

From a quantum mechanical perspective, this apparent emergence of
classicality, namely how and when an object loses its quantum
features and becomes indistinguishable from a classical
description, can be explained to a large extend by decoherence
theory \cite{Joos2003a,Schlosshauer2007a,Hornberger2007a}. The
crucial point is to acknowledge that no quantum object is
completely isolated. It is embedded in an environment consisting
of gas particles, photons and the like. Since the environmental
state gets very quickly correlated due to quantum interactions
with the object,  information about the whereabouts of the
quantum object is rapidly disseminated into the surroundings. An
initially pure state of the molecule is thus quickly replaced by
a mixed one, once the environmental  state is disregarded due to
its practical inaccessibility. The absence of quantum behavior in
the macroworld is then a natural consequence of the fact that
bigger and more complex objects are much harder to isolate. The
particular features of the environmental interaction and the
resulting transfer of which-path information thus lead to the
emergence of classical behavior in the motion of mesoscopic quantum
objects.

What are possible effects that might destroy a molecule's
coherent, delocalized state? One can conceive all sorts of scattering
processes with massive particles and light, originating from
local sources such as residual background gases and thermal
photons up to neutrinos and cosmic radiation. In our experiments, there are
at least two relevant and ubiquitous mechanisms that serve to
effectively measure the position of a molecule.
The first is due to collisions with background gas molecules,
while the second involves radiation emitted by the thermally
excited diffracting molecules.

\begin{figure}[tbp]
\begin{center} \includegraphics[width=1\columnwidth]{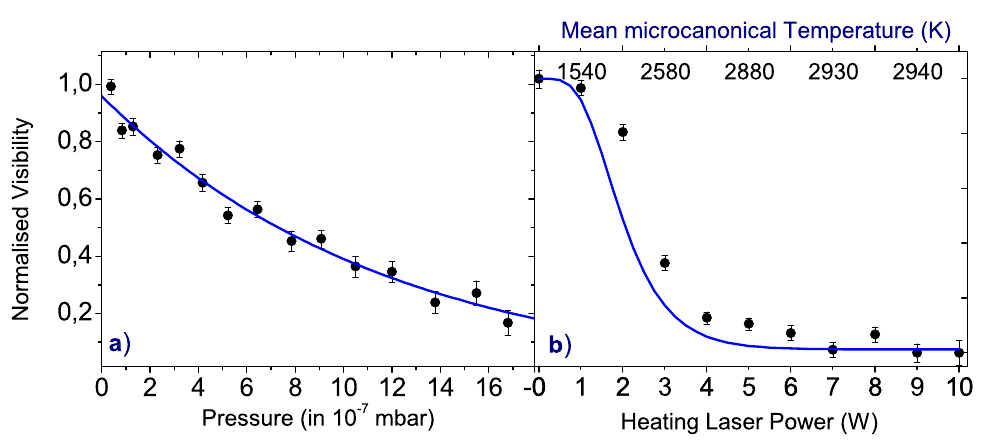}
\caption{Left: Collisional decoherence leads to an  exponential
decay of the fringe visibility with increasing residual gas
(methane) pressure~\cite{Hornberger2003a}. This observation is in
good agreement with the scattering theory calculation (solid
line). Right: Heating the fullerenes with a  laser prior to entering
the interferometer leads to a non-exponential decay of the
interference visibility. It is quantitatively explained by the
heat radiation due to the molecular temperature (upper scale),
which in turn is extracted from the heating dependence of the
detection efficiency~\cite{Hackermuller2004a,Hornberger2004a}.}
\label{fig7}
\end{center}
\end{figure}

\subsection{Collisional decoherence}

To find out how the interaction with a background gas can destroy
the interference pattern and lead to classical behavior, we
gradually added various gases to the vacuum chamber of our
Talbot-Lau interferometer during the experiments with C$_{70}$
fullerene molecules, see Fig.~\ref{fig6}a. We found that the
amount of contrast between the interference fringes fell
exponentially as more gas was added, and that the fringes
disappeared almost entirely when the pressure had reached just
$10^{-6}$\,mbar \cite{Hornberger2003a}. This can be seen in
Fig.~\ref{fig7}a for the case of methane gas. In fact, this
observation is in full quantitative agreement with a theoretical
analysis of the scattering processes
\cite{Hornberger2003b,Hackermuller2003b}. Although a single
collision with a gas molecule will not kick the massive fullerene
out of the interferometer path, it is enough to destroy the
interference pattern. The effect can be  explained by the
momentum exchange experienced, or equivalently, by the fact that
the scattered gas particle carries sufficient information to
determine the path that the interfering molecule has taken.

Since the gas particles at room temperature
have large  velocities, centered around $\tilde{v}_\text{g}=\sqrt{2k_\text{B}T/m_\text{g}}$,
it is justified to assume that a beam molecule gets completely localized by a single collision.
The exponential decay is thus directly related to the collision
probability, described by the thermally averaged, effective cross section \cite{Hornberger2003a,Hackermuller2003b,Vacchini2004a}
\[
\sigma_\text{eff}(v_m)=\frac{4\pi\Gamma(9/10)}{5\sin(\pi/5)}
\left(\frac{3\pi C_6}{2\hbar}\right)^{2/5}\frac{\tilde{v}_\text{g}^{3/5}}{v_\text{m}}
\left[1+\frac{1}{5}\left(\frac{v_\text{m}}{\tilde{v}_\text{g}}\right)^2\right].
\]
Here, $v_\text{m}$ is the velocity of the beam molecules, the
constant $C_6$ describes the van der Waals potential
$U(r)=-C_6/r^6$ between molecule and gas particle, and $\Gamma$
denotes the Gamma function. For a TLI with gratings  equally
separated by the distance $L$ the suppression of the visibility
at gas pressure $p_\text{g}$ is then given by
\[
\frac{V(p)}{V_0}=\exp\left(- \frac{2L \sigma_\text{eff}(v_m)}{k_\text{B}T}p_\text{g}\right).
\]
The observed contrast reduction was in good agreement with this
formula for the various gases we investigated, see
Fig.~\ref{fig7}a. Our calculations suggest that molecules  as
massive as 10$^6$\,amu would  still be unaffected by collisional
decoherence in a realistic Talbot-Lau interferometer, provided
the pressure does not exceed 10$^{-10}$\,mbar, which is not
trivial but certainly feasible with existing vacuum technologies.

We note that the concept of an index of refraction, which was
discussed in Sect.~\ref{sec:elementary} can be extended to the
case where a background gas acts as a medium for the matter waves.
One cannot describe collisional decoherence this way, but it may
still be used to account for the coherent modification due to the
background medium, as well as for the dampening of the beam due
to collisions leading to a complete loss. This gas-induced index
of refraction $n$  is determined by the forward scattering
amplitude $f_0$ as $n=1+\hbar^2\pi n_\text{gas}(\mb{r})\langle
f_0\rangle/(m_* E)$, with $n_\text{gas}(\mb{r})$  the local gas
density and $m_*$ the reduced mass; it involves an appropriate
averaging over the thermal velocity distribution in the gas
\cite{Champenois2007a,Hornberger2008a}. According to the optical
theorem $f_0$ has an imaginary part determined by the total
scattering cross section. It renders $n$ complex and thus
describes the dampening of the matter wave beam due to
collisional losses. The phase shifts and losses described by this
index of refraction have recently been measured with an atomic
Mach-Zehnder interferometer~\cite{Schmiedmayer1995a,Jacquey2007a}.

\subsection{Thermal decoherence}

We now ask how the `internal temperature' of a molecule affects
its ability to  interfere. The concept of an internal temperature
is not particularly meaningful for atoms, not to speak of
elementary particles. However, for  complex objects such as large
molecules it is very natural to describe the energy distribution
of the many vibrational and rotational degrees of freedom in terms
of the micro-canonical temperature.

This applies in particular to fullerenes, which are very stable
and which are able to store energies corresponding to an internal
temperature of up to 5000\,K before they start decomposing in free
flight. Heating can be done using intense visible laser light.
Each green photon of an Argon ion laser (514\,nm) increases the
average internal temperature by about 170\,K and the setup
allowed to deposit up to about 50 photons per molecule. 

Hot fullerenes are known to emit heat radiation in a continuous
spectrum similar to a black body~\cite{Kolodney1995a}. One must,
however, account for the frequency dependence of the emission
cross section, rendering fullerenes 'grey'  rather than black
bodies. In addition, their finite heat capacity plays a role as
well as the fact that the emission is not in thermal equilibrium
with the external thermal radiation field \cite{Hornberger2005a}.
Also, competitive processes such as electron and C$_2$ fragment
emission have to be taken into account.

\begin{figure}[tbp]
\begin{center}
\includegraphics[width=0.6\columnwidth]{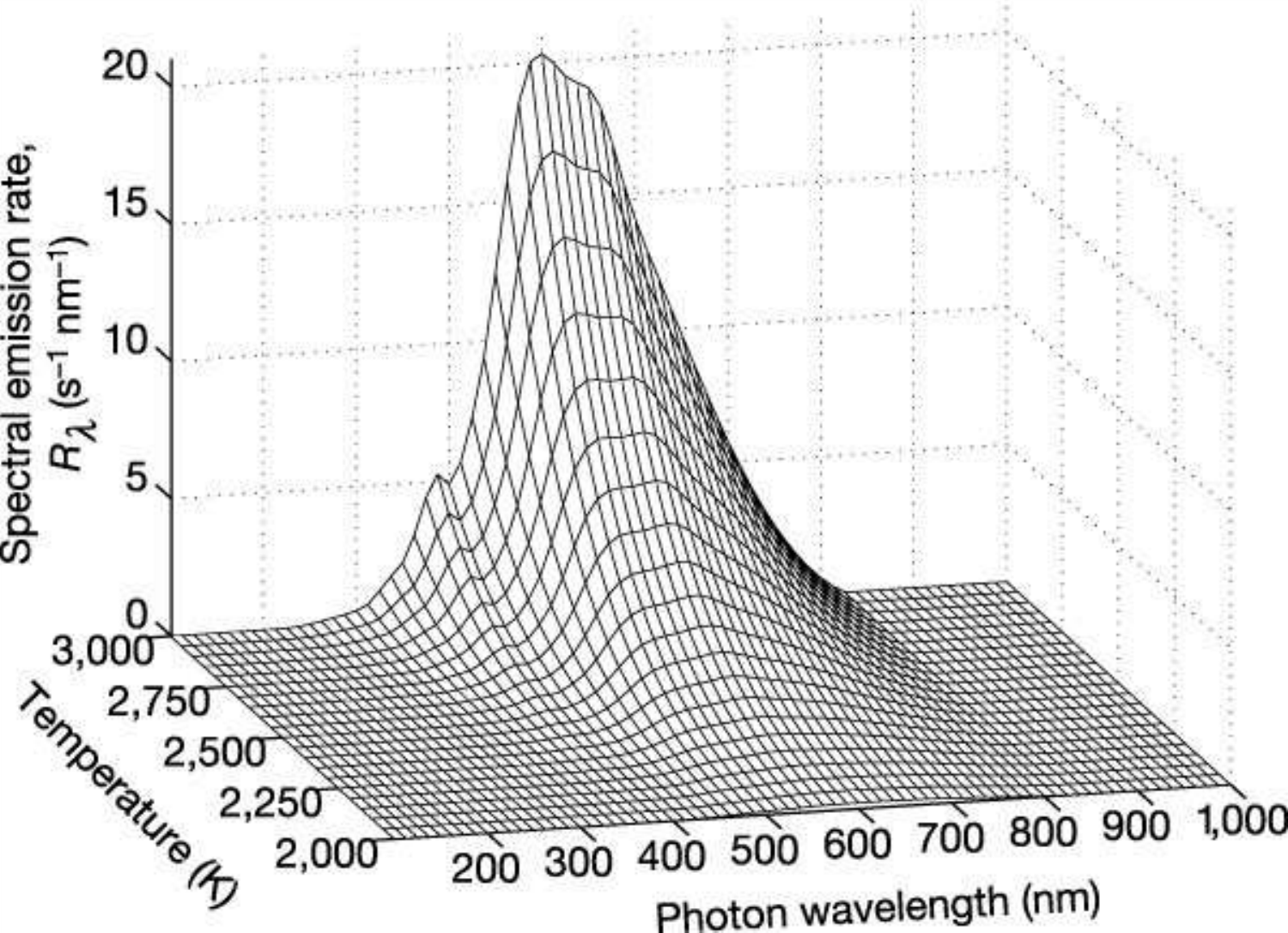}
\caption{Spectral photon emission rate of C$_{70}$ fullerenes as a function of their internal (microcanonical) temperature \cite{Hackermuller2004a}. It is used to calculate the decohering effect of thermal heat radiation. The suppression of long wave length photons is due to the electronic excitation gap in C$_{70}$  of about 1.5\,eV.} \label{fig8}
\end{center}
\end{figure}

Based on a correspondingly adapted version of Planck's law, which involves the
measured frequency dependent absorption cross section, one can
describe the spectral photonic emission rate $R_\lambda$ as a function of the
internal molecular temperature, see Fig.~\ref{fig8}. According to  decoherence theory,
all it takes to destroy the interference fringes is for the
molecule to emit either lots of long-wavelength photons or a
single photon with a wavelength shorter than about twice the
separation between the coherently split molecular wavelets.

This can be seen from the formula for the visibility suppression,
which involves an integration over all photon wavelengths
$\lambda$ and over all longitudinal positions $z$ in the
interferometer \cite{Hackermuller2004a,Hornberger2005a},
\[
\frac{V(T)}{V_0}=\exp\left[
-\int_0^{2L}\frac{\mathrm{d}z}{v_z}\int_0^\infty\mathrm{d}\lambda\, R_\lambda(\lambda,T)
\left\{1-
\mathrm{sinc}\left(2\pi\frac{d}{\lambda}\frac{L-|z-L|}{L_\text{T}}
\right)\right\}\right].
\]
At each position $z$ the argument of the function
$\mathrm{sinc}(x)=\sin(x)/x$ compares the photon wavelength to
the effective path separation, which varies (for $L=L_\text{T}$)
between $0$ and the grating period $d$. Whenever the photon
wavelength can `resolve' this path separation, which is largest at
the second grating,  the sinc function is substantially smaller
than unity and thus should give rise to a suppressed visibility.

Indeed, when we increased the internal temperature of C$_{70}$ fullerene
molecules to above 1000\,K, the contrast between the interference
fringes slowly disappeared, see Fig.~\ref{fig7}b
\cite{Hackermuller2004a}. In the experiment we varied the
temperature by heating the molecular beam in front of the
interferometer with a crossing laser beam, as shown in
Fig.~\ref{fig6}b. Our understanding of the beam temperature and
its cooling dynamics was independently checked by recording
the heating dependence of the detection efficiency at various
beam velocities \cite{Hornberger2005a}. Using this temperature
variation as an input to our microscopic decoherence model
the observed decoherence rates were reproduced quantitatively,
as can be seen  by the solid
line in Fig.~\ref{fig7}b. This good agreement between the
predicted and the measured decoherence rate indicates that the
C$_{70}$ molecules emitted a few visible photons (with a broad
band  of wave lengths centered around 800\,nm) when they were
heated to internal temperatures above 2500\,K. Since the grating
slit separation was 1\,$\mu$m, this sufficed to substantially
reduce the fringe visibility, as enough `which-path' information
became available to the environment.

This experiment proves that decoherence due to heat radiation can
be quantitatively traced and understood. It confirms the view
that decoherence is caused by the flow of information into the
environment, mediated by a transfer of momentum. Finally, it
shows that thermal decoherence is very relevant for mesoscopic
and macroscopic objects. Large molecules are already sufficiently
complex to serve as their own internal heat bath, leading to
auto-decoherence, provided they are sufficiently hot. In that
sense, they can be regarded as condensed matter systems.

Fortunately, this effect will be less of a concern in future
interferometry experiments with large molecules, clusters or
nanocrystals. Objects like these will have to be substantially
cooled in their external degrees of freedom, to make them
coherent in the first place. All cryogenic cooling methods will
have to additionally reduce the internal temperature and thus
the probability for the emission of thermal radiation.

\subsection{Dephasing and phase averaging}

In addition to the proper decoherence mechanisms mentioned above,
we have to deal with a number of other issues that do not involve
the quantum interaction with the environment, but which may still
significantly affect the interference quality. In practice, these
phenomena are hard to distinguish from genuine decoherence, even
though their theoretical description is rather different.

A simple, yet very relevant effect is the phase averaging brought
about by acoustic vibrations of the interferometer. Due to these
vibrations subsequent molecules 'see' the gratings at slightly
different spatial locations, so that each interfering molecule
gives rise to a probability distribution that is out of phase
with respect to a previous one. The resulting blurring of the
observed  interference pattern can be quite
strong~\cite{Stibor2005a}. At certain frequencies vibration
amplitudes as small as 15\,nm can be detrimental, even if the
grating constant is as wide as 1\,$\mu$m.   This finding imposes
severe constraints on the allowed vibration amplitudes for all
future experiments, which will inevitably have to work at even
smaller grating constants. A related issue is the thermal drift
in the interferometer apparatus, which must be eliminated or
controlled with a similar degree of accuracy.

Another, equally coherent effect, that may affect the
interference contrast,  is due to the fictitious forces brought
about by the fact that the laboratory is not an inertial system.
In the laboratory frame the molecules are accelerated by the
gravitational field and they experience a Coriolis force due to
Earth's rotation. Matter wave interferometers are very sensitive
to these effects, and in fact, atom interferometers are currently
among the best devices for measuring the corresponding
accelerations~\cite{Ekstrom1995a,Gustavson1997a,Peters1999a}.

If the molecular beam was monochromatic there would be no
reduction of visibility, only a shift of the pattern due to the
non-inertial forces. However, this shift is at least inversely
proportional to the beam velocity~\cite{Stibor2005a}. Even though
the interferometer would accept a rather broad velocity
distribution if it was at rest in an inertial system, the
fictitious forces now introduce a strong dispersion by shifting
the interference pattern to different positions for different
velocity components. This introduces again an effective blurring
of the interference signal which is aggravated for increasingly
massive particles with small beam velocities.

In particular, in case of a constant acceleration $a$ the interference
contrast gets reduced by
\begin{equation*}
  \frac{V}{V_0}= \exp \left( -2 \left[ \pi \frac{a L^2 \sigma_v}{d\, v_z^3}\right ]^2\right)
\end{equation*}
where $d$ is the grating period, $v_z$ is the longitudinal velocity,
$\sigma_v$ its spread, and $L$ is the grating separation.
In the case of gravity, we can substitute $a= g\, \sin\theta$, where $\theta$ gives the tilt of the interferometer plane with regard to the
local direction of the gravitational acceleration.
For the Coriolis force we may substitute $a = 2 v \, \Omega_0 \,
\cos \phi$, where  $\phi$ is the angle between the normal on the interferometer plane and the rotation axis
of the earth, spinning at angular frequency $\Omega_0$.

\section{Matter wave interferometry for molecule metrology}
\begin{figure}[tbp]
\begin{center}
\includegraphics[width=\columnwidth]{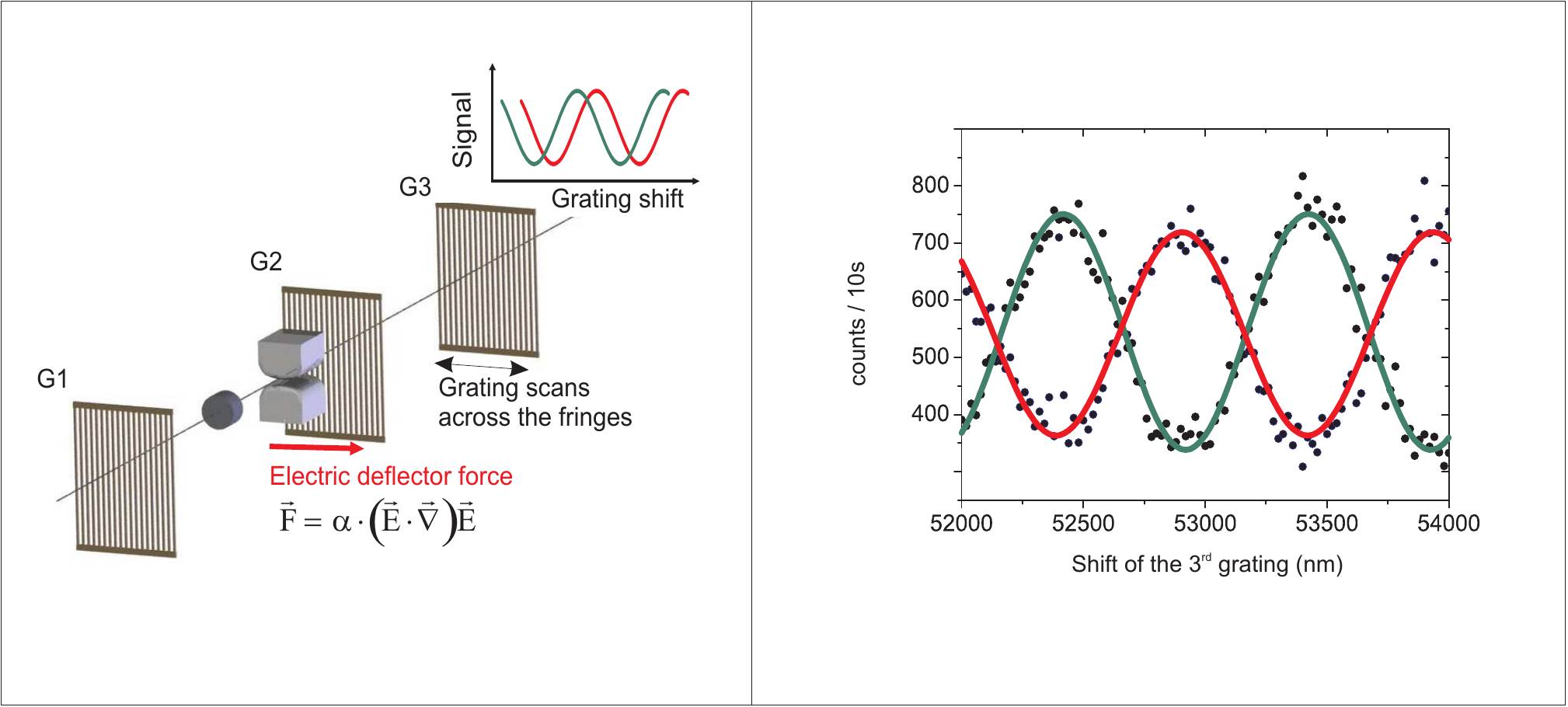}
\caption{left: Setup for interferometric deflectometry.
The traversing clusters and molecules interact with both the
optical phase grating and the electric field gradient via their
polarizability and/or electric dipole moment. Generalizations to
measurements of magnetic moments and magnetic susceptibilities are well
conceivable. right: The fringe shift of the molecular
interferogram in the external electric field permits to  directly
determine the molecular polarizability with high
precision. This curve was obtained with three mechanical
gratings, i.e., in the Talbot Lau setup [95].} \label{fig9}
\end{center}
\end{figure}

Let us now turn to the question how the interference of molecules
can be used to measure molecular properties. A first application,
which was demonstrated recently, is the measurements of the
static and the  dynamic polarizability of large
molecules~\cite{Berninger2007a, Hackermuller2007a}.

In contrast to earlier atom interferometric polarization
measurements~\cite{Chapman1995b}, the  Talbot-Lau concept does
not allow one to address two individual interference paths
separately. However, the gratings  imprint a nanostructure onto
the molecular beam, which admits the measurement of tiny beam
shifts with high precision. At the same time, the interaction
between the flying molecules and both the material and the
optical gratings depends sensitively on the electric
polarizability, or more generally the electric susceptibility,
which also includes the molecular permanent electric dipole
moments.

The experiment shown in Fig.~\ref{fig9} depicts a Talbot-Lau
interferometer into which a specifically
designed deflection electrode was inserted~\cite{Stefanov2008a}. The electrode
is designed to provide a most homogeneous electric force field $F
=\alpha (\vec{E} \vec{\nabla})\vec{E} $ over the cross section of
the laterally extended molecular beam. This leads to a molecular
beam deflection  at the location of the third grating
proportional to
\begin{equation*}
  x_d \propto \frac{\alpha}{m} \frac{(\vec{E}
  \vec{\nabla})\vec{E}}{v_z^2}.
\end{equation*}
When we vary the voltage $U$, and thus the electric field $E$
across the electrodes, we find a shift of the molecular
interference pattern. By fitting a parabola to the data we can
extract the polarizability with an accuracy of currently about 4
percent. This good degree of precision results from the high
resolution of the fringe shift of 10...15\,nm, and from the velocity
spread in the present beam configuration, which amounts to about
15\%.

These parameters can certainly still be improved. Future
deflectometer measurements will clearly profit from molecular
beams with narrower velocity bands. On the other hand, it is worth
noting that even at the present resolution interesting
statements can be made about the internal molecular state.
Molecules of identical mass and composition may have dramatically
different structures and conformations giving rise to strongly
different electrical response functions. This is the case,  in
particular, for sequence isomers of polypeptides and for
conformation isomers of azobenzene derivatives or retinal.
Similarly, metal clusters of the same mass may be in different
magnetic states. It is thus conceivable to analyze such isomers,
and even to sort mixtures of them, by using a Talbot Lau
interferometer ~\cite{Ulbricht2008a}.

The Kapitza-Dirac-Talbot-Lau interferometer described in
Sect.~\ref{sec:TL} offers the possibility to also measure the
{\em optical } polarizability at the frequency of the standing
light field \cite{Hackermuller2007a}. One records the interference
visibility as a function of the laser power and the mean velocity
in the beam. By fitting the results of the quantum calculation to
these data, one can extract the complex dynamic polarizability
with good accuracy, similar to the mentioned deflection
experiment.

One may also think of combining the deflectometry scheme with the
Kapitza-Dirac-Talbot-Lau interferometer. This way the
contribution of permanent electric dipole moments could be
accessed, since the optical field changes its sign at a frequency
way too fast for the dipoles to follow, while the static field
may orient them. A KDTLI-deflectometer thus opens the possibility
to measure the static susceptibility and the optical
polarizability at the same time. A direct comparison should then
allow one to separate the influence of polarizability and static
dipole moments and to determine them independently with good
accuracy in future experiments.

These investigations will become increasingly relevant as new
methods are being developed to cool the internal molecular states
as well. Molecule interferometry may then become an interesting
add-on to mass spectrometry~\cite{Gerlich2008a}.

\section{Optical imaging of sub-wavelength molecular nano-interferograms}
\begin{figure}[tbp]
\begin{center}
\includegraphics[width=0.6\columnwidth]{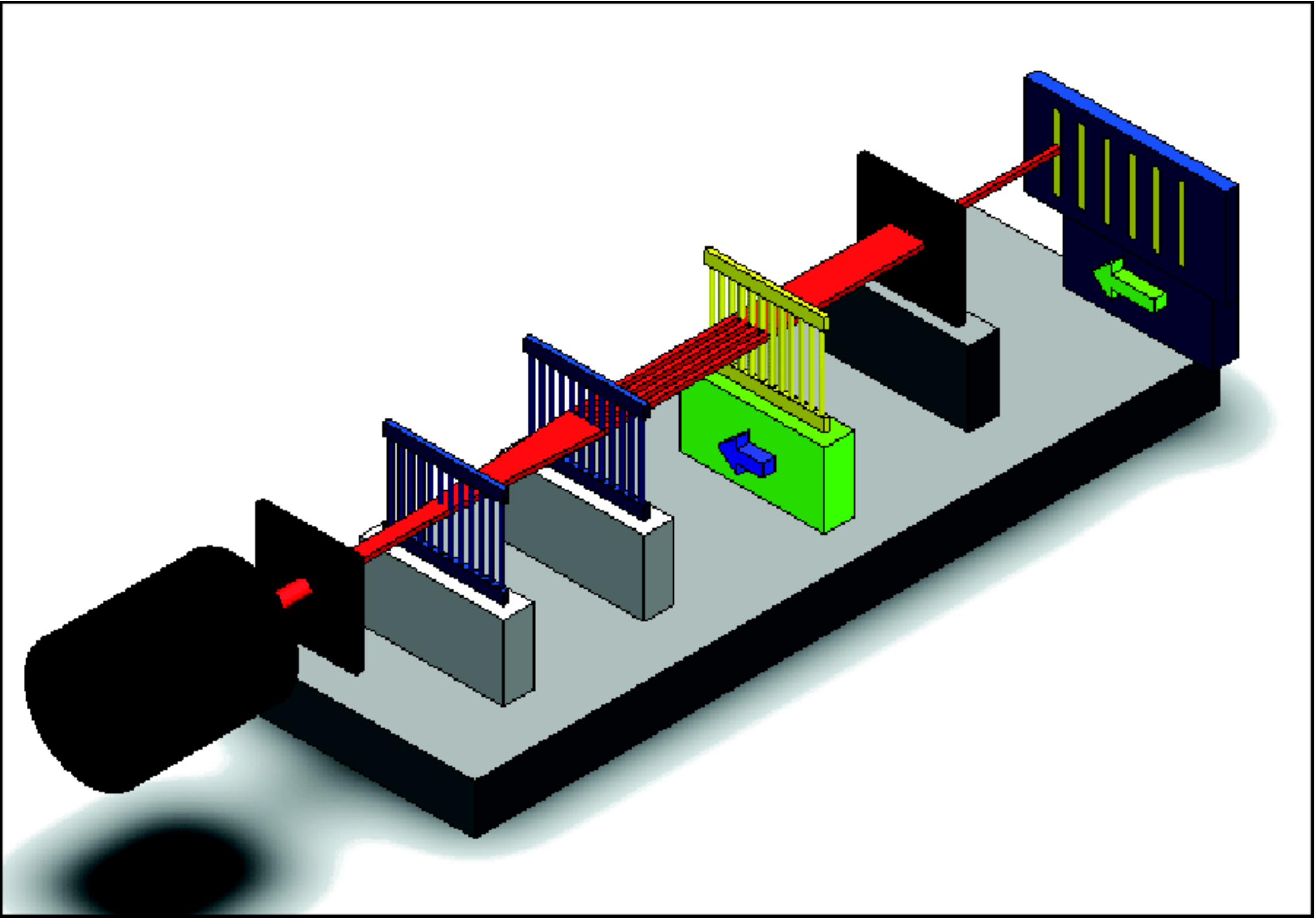}
\caption{Setup for mechanically magnified imaging of a
near-field molecular interferogram. The Talbot-Lau interferometer
transmits a molecular beam over a restricted lateral width of
about 100$\,\mu$m. This beam is collected on a quartz surface
after mechanical magnification as described in the text. The
longitudinal velocity is encoded in the falling height, i.e., the
{$y$-position} of the molecular beam on the screen. The position of
the third grating is encoded in the $x$-position of the comoving
quartz plate~\cite{Stibor2005b}.} \label{fig10}
\end{center}
\end{figure}

In the previous chapters we focused primarily on the methods
required to prepare and maintain coherence, and to use it in interferometric measurements. However, daily life in the lab is also strongly concerned with the
preparation of sufficiently intense molecular beams, and with methods to efficiently detect the slow, neutral  molecules.

It is beyond the scope of this contribution to
summarize all methods that were explored and developed in the
context of our Vienna experiments on molecule interferometry.
One  example seems particularly instructive, though, and
useful also in future experiments.
It is the mechanical
magnification of surface-recorded molecular interferograms,
which serves as an alternative detection scheme.

The ionization of organic molecules with masses  beyond a few
thousand amu is generally considered to be a big challenge. Often
the ionization energies are too high to be accessible with
available laser light, or  neutral fragmentation is energetically favored~\cite{Becker1995a}. Efficient photo
detection  of neutral organic clusters  is therefore only
possible in selected cases, such as for
Trp$_{30}$Ca~\cite{Marksteiner2008a}.

On the other hand, large organic molecules often fluoresce quite
efficiently. Even if they cannot do this natively they can be
labeled with fluorescent dye tags. In this sense, fluorescence
detection gets increasingly simple for large molecules. One might
therefore consider the direct optical imaging of interferograms.
An obvious problem with this is that the  quantum interference
fringes of high-mass molecules will  generally be rather narrow,
in fact often below Abb\'{e}`s optical diffraction limit $a\simeq
\lambda/2$.

A simple, yet very efficient approach to the imaging of periodic
sub-wavelength molecular structures is the use of mechanical
magnification. In the experiment we enlarge the interference
pattern by about 1:4000 in order to facilitate the imaging of the
interferogram. An illustration of the setup, shown in
Fig.~\ref{fig10}, explains the idea: we place a transparent
recording quartz plate behind the mechanical TLI and thus collect
all molecules that pass the interferometer. The counting of the
plate-deposited molecules serves only one purpose: measuring the
transmitted molecular flux. It may in addition happen that a
fine-grained molecular interference pattern also forms on the
quartz plate if the plate is positioned in the proper distance
behind the third grating. However, we do not use this
information. In the end we only record the total fluorescent
light, which is proportional to the integrated number of
transmitted molecules, as long as the deposited layer is
sufficiently thin.

\begin{figure}[tbp]
\begin{center}
\includegraphics[width=0.8\columnwidth]{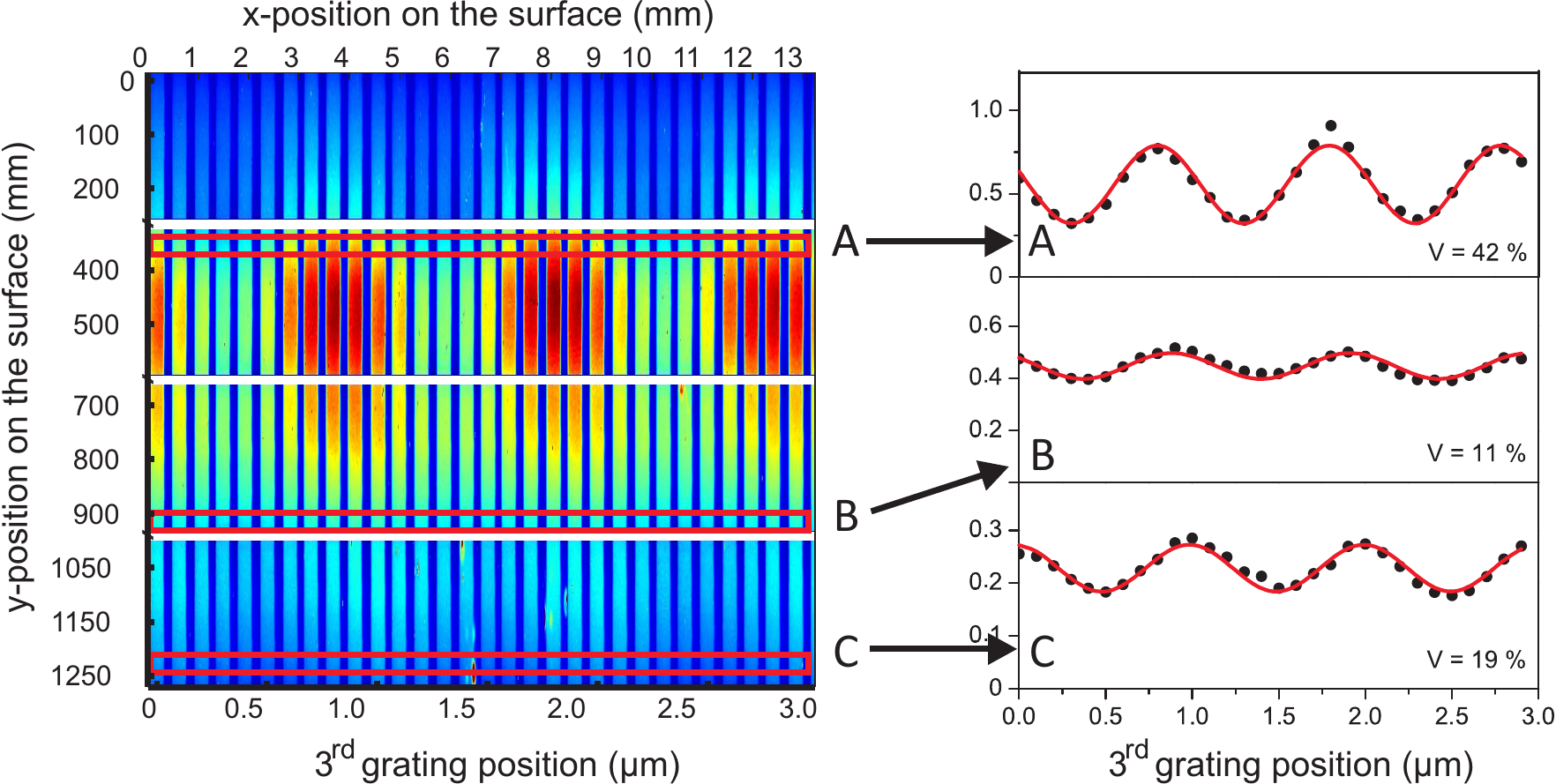}
\caption{Left:  Fluorescence microscopy image of a porphyrin
interferogram with mechanical magnification. One clearly observes
interference patterns of different contrast for different
molecular velocities. Right: A trace along three selected vertical
positions reveals the interference fringe more
clearly~\cite{Stibor2005a}.} \label{fig11}
\end{center}
\end{figure}

The interference information is now magnified and encoded
by scanning the third grating across the molecular density pattern as in the Talbot-Lau experiments described in Sect.~\ref{sec:TL}.
For each discrete scan position of the third grating (step size 100\,nm)
the quartz plate is moved by a significantly larger distance, for instance
400\,$\mu$m, corresponding to a 4000-fold mechanical
magnification.  The resulting large area molecular deposit can
then be read in fluorescence microscopy. The result of this is
presented in Fig.~\ref{fig11}. The abscissa encodes the transverse position of the third grating, while the ordinate encodes the vertical position of the molecules in the lab frame. Here we explicitly make use of
a gravitational velocity selection scheme~\cite{Nairz2001a}. It
exploits that slow molecules have a longer time of flight. They fall deeper and thus arrive lower on the plate than the fast molecules.
The optical recording thus permits us  to directly reveal the
functional relation between the fringe visibility and the
molecular de Broglie wavelength (velocity), as shown in
Fig.~\ref{fig11}b.

This trick of optically imaging sub-wavelength molecular interferograms by means of mechanical magnification is a scalable method.  It may be applied to particles of
any mass, as long as they fit through the grating slits and as long as they can be marked by fluorescent dyes.

\section{The future of quantum experiments with clusters and molecules}

Our experience in the realm of quantum experiments with complex
compounds taught us that sizable challenges are piling
up if one tries to increase the mass, size and complexity of
the interfering object. At the same time, many interesting
avenues are opening up:

On the one hand, it remains an important goal to explore the
ultimate mass limit of quantum interferometry. This is a worthy
and thrilling experimental challenge in itself. In addition, as
mentioned in Sect.~\ref{LargeMolecules}, various theoretical
proposals have emerged in recent years, some of them still in a
rather early stage of formal justification, that suggest that
fundamental effects might suppress the visibility of quantum
interference beyond a certain mass limit. This implies the
experimentally appealing perspective, that the persistence of
quantum interference beyond a certain mass and coherence time may
allow us to test and potentially falsify certain unconventional
extensions of quantum mechanics and theories of quantized space time. The
current range of parameters is so wide that some models start
being touched by our current experiments, while others  still
lie  in the future by six to seven orders of magnitude in mass.
That is pretty far---but not beyond the means of feasibility if
the required reasonable experimental resources can be focused on
this project.

Even on a much more modest mass level, in the range between 500
and 10.000 amu, there is a plethora of molecules that only wait
for being subjected to quantum interferometry. We have already
seen that matter interferometry allows us to determine static and
optical polarizabilities with good precision. Already these
scalar values provide relevant information that may help
identifying or separating~\cite{Ulbricht2008a} molecular
conformers of different amino acid or nucleotide sequences, i.e.
of polypeptides, oligonucleotides or short DNA strands.

Structural and conformation changes in free flight may become
measurable. An example for that would be the trans-cis
isomerization in azobenzene derivatives or retinal. Their
different conformers are connected with different polarizabilities
and dipole moments and thus affect the highly sensitive molecule
grating interaction. Interferometry may thus become an
interesting complementary tool to mass spectrometry and optical
spectroscopy.

The Vienna experiments have proven that quantum interference is
feasible in spite of the substantial internal excitations present
in complex molecules. On the other hand, the investigations also
show that further slowing and cooling will be required in future
experiments to pave the path to interferometry with supermassive
clusters and highly complex molecules. Only internal state
cooling will allow us to play with new ideas on internal-external
state entanglement. Highly efficient cryogenic techniques will
probably have to be complemented by new quantum optical
manipulation ideas.

In variance of Feynman's early statement on
nanotechnology~\cite{Feynman1959a} we therefore think: {\em There
is plenty of room at the top}.

\acknowledgments The scientific program on macromolecular
interference was  started in collaboration between one of the
authors (MA) and Anton Zeilinger. Over the years, work on
interferometry with fullerenes and large molecules has benefited
enormously from important contributions by many bright students,
postdocs and collaborators, as documented in the references. Our
research has been supported by the Austrian FWF within the
programs SFB F1505, START Y177, Wittgenstein Z149 and COQUS, 
by the European
Commission within the IHP network QUACS and the ESF programme
EuroQuasar MIME, and by the DFG Emmy Noether programme.


\end{document}